\documentclass[amsmath,amssymb,reprint,bibnotes]{revtex4-2}
\usepackage{graphicx}
\usepackage{bm}
\usepackage{amsmath} 
\usepackage{graphics}
\usepackage[dvipsnames]{xcolor}
\usepackage[mathlines]{lineno}
\usepackage{braket}
\usepackage{subfigure}
\usepackage[compact]{titlesec}
\begin{document}
\title{Stark Control of Plexcitonic States in Incoherent Quantum Systems}
\author{Hira Asif$^{\bf (1)}$}
\author{Ramazan Sahin$^{\bf (1)}$}
\email{rsahin@itu.edu.tr}

\affiliation{${\bf (1)}$ {Department of Physics, Akdeniz University, 07058 Antalya, Turkey}}

\date{\today}

\begin{abstract}
Electro-optic control of quantum dots embedded in the plasmonic nanocavities enables active tuning of photonic devices for emerging applications in Quantum optics such as quantum information processing, entanglement and ultrafast optical switching. Here, we demonstrate the coherent control of plexcitonic states in (i) an off-resonant and (ii) a resonant coupled quantum systems through optical Stark effect (OSE). We analyze a hybrid plasmon-emitter system which exhibits tunable Fano resonance, Stark induced transparency (SIT) and vacuum Rabi splitting due to quadratic Stark shift in the degenerate states of quantum emitter (QE). In addition, a resonantly coupled system shows the signature of double Fano resonance due to Stark-induced splitting in a two-level QE. Our study shows that Stark tuning of plexcitons not only mitigates decoherence in the quantum system but it also stimulates on/off switching of spontaneous photon emission in the visible regime. Such tunable systems can be used to operate photonic integrated circuits (PIC) for applications in quantum computing and information processing.
\end{abstract}

\maketitle

\section{INTRODUCTION}
Active control of quantum states in an incoherent system has become a new challenge for operating multifunctional in-situ programmable photonic integrated circuits (PIC) \cite{Fang2015, Giordani2023, Wang2019}. Controlling these systems in real-time and obtaining desired properties is crucial for the development of quantum technologies such as quantum information processing, quantum computing and single photon sources \cite{Imamoglu1999,Kim2020}. In this regard, Quantum Plasmonics provides a fastest route to achieve such control by manipulating quantum properties of surface plasmons and excitons such as quantum emitter (QE) through cavity quantum electrodynamics (CQED) \cite{Tame2013, Vasa2018}. Due to its extreme field confinement, plasmon cavity entails profound coupling with QE which enables coherent control of quantum devices at a single photon level \cite{Toma2015,Vasa2018}.\\
A coherent interaction of plasmon cavity with exciton generates mixed quantum states also known as plexcitonic dressed states which inherently possess all the information of controlled quantum system \cite{Fofang2008}. Tuning these states as a function of coupling strength yields two distinct phenomena i.e. Fano Resonance (FR) and vacuum Rabi Splitting (RS) for intermediate and strong coupling regimes, respectively \cite{Leng2018, Liu2017,Toma2015}. Both coupling mechanism enable coherent oscillations and allows quantum superposition of different states which are essentially important for entanglement and quantum information \cite{emre2022, Lorenzo2017}. So far FR and RS have been demonstrated in the resonantly coupled plasmon-emitter systems while coherent control has been achieved by modulating the geometrical parameters of the structure \cite{Miroshnichenko2010}, and this demands challenging fabrications process.\\
Nevertheless, achieving coherent control in the off-resonant coupled quantum system is of potential importance for implementing active photonic functionalities like ultrafast switching \cite{Schwartz2011}, signal processing and lasing at nanoscale \cite{Oulton2009,Klimov2007}. One of the promising way to manifest dynamical control over quantum systems is to use the external variable influence. At present active tuning of polaritonic modes under CQED treatment has been explored either through resonant excitation \cite{Vasa2013}, magnetic tuning \cite{Li2021}, dielectric control 
\cite{Hapuarachchi2017} or by exploiting voltage tunable 2D materials \cite{Amin2013} and quantum dots \cite{Asif2024}. However, coherent tuning of off-resonant quantum states in real-time through Optical Stark Effect (OSE) has not been discussed yet to the best of our knowledge.
\begin{figure}
\includegraphics[scale=0.5]{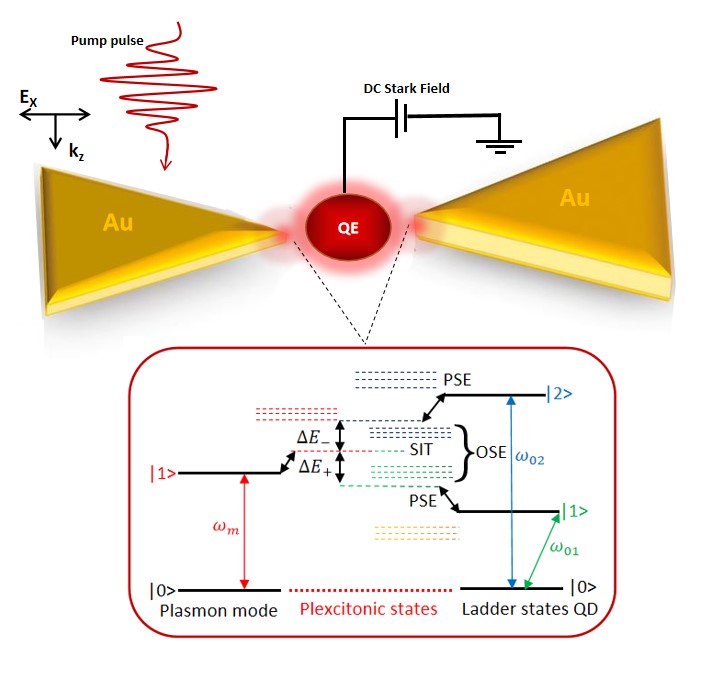}
\caption{\label{fig:1} Schematic diagram of a hybrid system of Au bow-tie nanoantenna and a three-level QE. QE is attached to external voltage-bias which induces Stark shifts in the hybrid plexcitonic states.}
\end{figure}
\begin{figure*}[t!]
    \centering
    \begin{subfigure}
        \centering
        \includegraphics[height=2in]{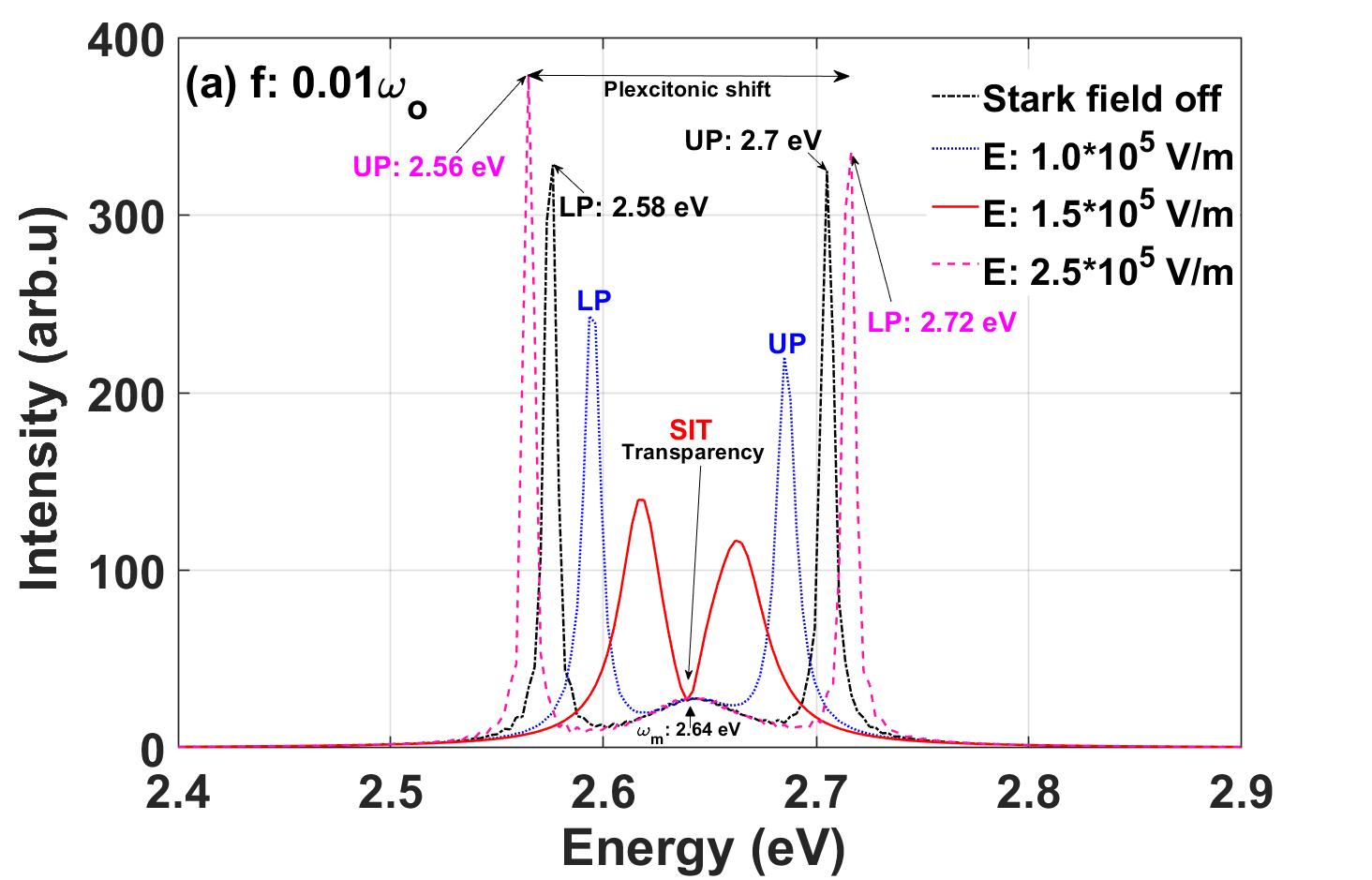}
    \end{subfigure}%
    ~ 
    \begin{subfigure}
        \centering
        \includegraphics[height=2in]{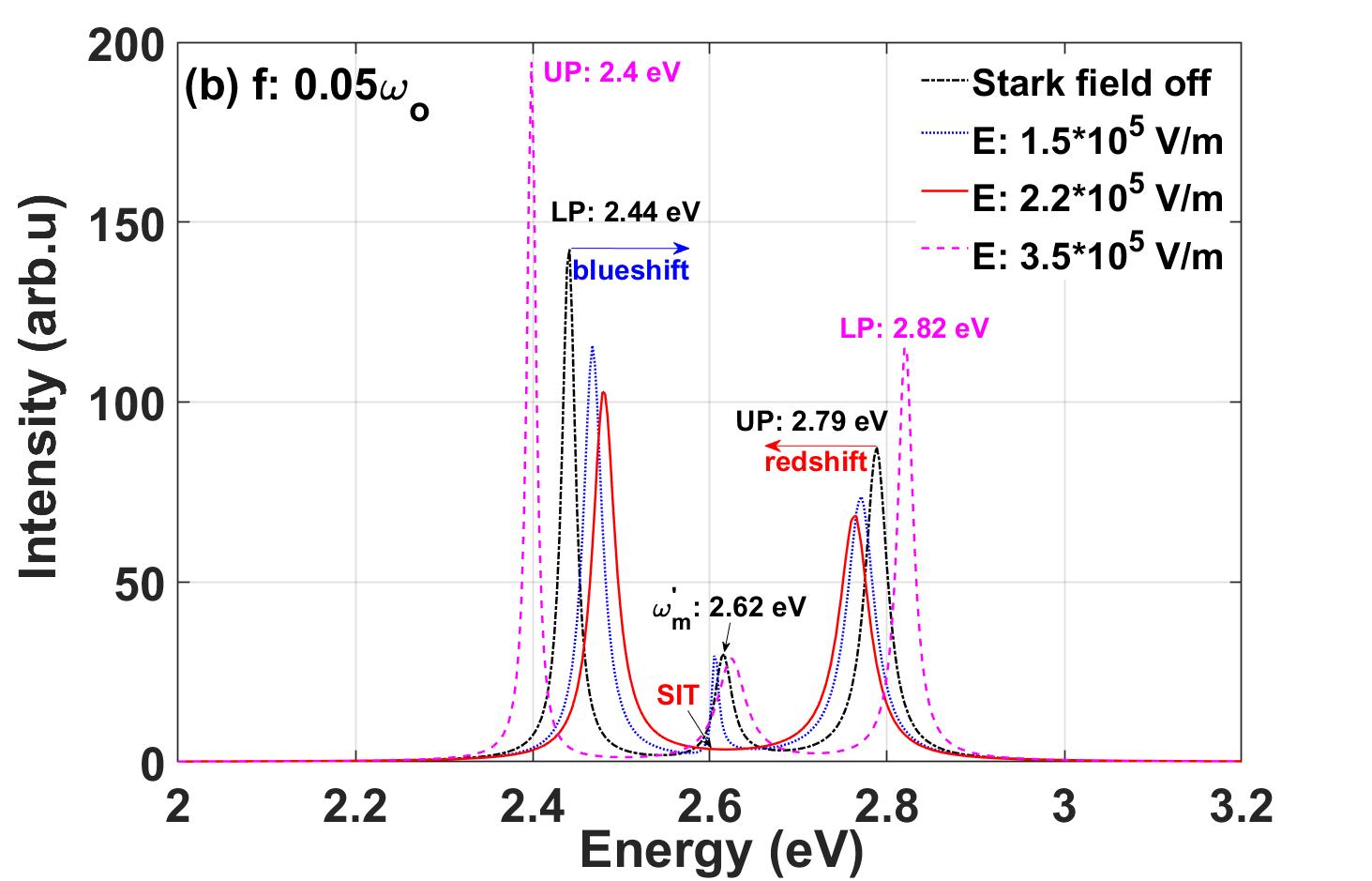}
    \end{subfigure}
    \caption{\label{fig:2} Stark induced shifts and transparency in the plexcitonic states (off-resonant coupling) for the (a) weak (f=0.01$\omega_o$) and (b) intermediate coupling strength (f=0.05$\omega_o$), respectively. Transparency window (SIT) appears at plasmon mode frequencies, (a) $\omega_m= 2.64$ eV and (b) $\omega^{'}_m= 2.62$ eV for an optimum value of Stark field. The eigen frequencies of the first and second excited states of QE are taken as $\omega_{01}= 2.58$ eV, $\omega_{02}= 2.7$ eV for the weak coupling case and $\omega_{01}= 2.5$ eV, $\omega_{02}= 2.7$ eV for intermediate coupling cases with decay rates $\gamma_{0j}=5 $ $\mu$eV, respectively.}
\end{figure*}         
In contrast to previous studies discussing Fano resonance and tunability of polariton modes through different plasmonic structures \cite{Chen2014,Cao2015,Lodewijks2013}, our work demonstrates an alternative approach of tuning plexcitonic modes of an incoherent quantum system via OSE. In this letter, we study Stark tuning of plexcitons in two scenarios. (i) In a hybrid plasmon-emitter system excited and coupled off resonantly, the probe (Stark) field shifts the eigenenergies of a three-level QE and coherently drives the off-resonant states close to resonance which leads to path interference (Fano resonance). The coherent phase shift of plexcitons generates a transparency window which we named as Stark Induced Transparency (SIT). Furthermore, we inspect how small perturbation in the Stark field yields large modulation in the vacuum Rabi splitting. (ii) In a resonantly coupled system, the Stark field lifts the degeneracy by splitting the excited state of a two-level QE which induces double Fano resonance in the system. Our concept of Stark tuning of hybrid modes explicate coherent control of quantum devices as a proof of principle concept which can be used to mitigate decoherence of a quantum system and also demonstrate active tuning of spontaneous photon emission in the visible regime.

\section{THEORETICAL FRAMEWORK}
We investigate the dynamics of a hybrid quantum system by formulating an analytical model that simply describes the interaction between a plasmon mode and QE in the context of a coupled harmonic oscillator \cite{Asif2022}. The hybrid plasmon-emitter system consists of Au bow-tie nanoantenna \cite{Sahin2019} and a QE, placed at distance R between nanodimmer, as shown in Fig.\ref{fig:1}. The bow-tie nanoantenna is irradiated with a p-polarized light of frequency ($\omega_{o}$) and amplitude ($E_o$), which creates intense localized dipole modes (LSP). We define ($\hbar\omega_m\hat{a}^\dagger\hat{a}$) as the unperturbed energy of LSP mode with annihilation $(\hat{a})$ and creation $(\hat{a}^\dagger)$ operators. The dipole interaction of the driving pulse with plasmon mode is expressed as $ \mathcal{M}= E_o \mu_m e^{-i\omega_o t}(\hat{a}^\dagger)+H.c$ \cite{Hapuarachchi2017},
where $\mu_m$ is the dipole matrix element define as $\mu_m=-i \sqrt{12\pi \epsilon_o \eta r^3 \hbar}$ with r as the edge size of bow-tie triangle. Although our methodological approach differs for two different cases (off-resonant or resonant coupling), the defined total Hamiltonian is mathematically the same. And hence, QE is specified as a three-level system in the ladder configuration with basis states expressed as raising $\hat{\sigma}^\dagger=\ket{e}\bra{g}$ and lowering $\hat{\sigma}=\ket{g}\bra{e}$ operators for both cases. The transition energies of $\ket{1}$ and $\ket{2}$ excited states of QE are defined as $\hbar\omega_{01}$, and $\hbar\omega_{02}$, respectively while the ground state energy is taken as zero. Since the pump-field is off-resonant to plasmon mode and QE, we assume that there is no interaction between the pump field and the QE for both cases. Therefore, we neglect the dipole coupling of QE. The interaction between plasmon mode and QE is quantified through a phenomenological constant $(f)$ measured as coupling strength between two oscillators. We choose the value of $(f)$ normalized to frequency of excitation field according to weak ($f= 0.01\omega_o$) and intermediate ($f= 0.05\omega_o$) coupling regimes as referenced as in \cite{Leng2018}. After applying rotating wave and dipole approximations \cite{Li2021}, we define the total Hamiltonian for a three-level QE interacting with plasmon mode \cite{Messina2003, Li1987} as follows,      
\begin{eqnarray}
\hat{\mathcal{H}}=\hbar\Delta_m\hat{a}^\dagger\hat{a}+\hbar(\Delta_{01}-\Delta E)\hat{\sigma}_{11}+\hbar(\Delta_{02}+\Delta E)\hat{\sigma}_{22}\nonumber\\
+i\hbar f\sum_{j}(\hat{a}^\dagger{\textdagger}\hat{\sigma}_{0j}-\hat{a}\hat{\sigma}_{0j}^\dagger)+\mathcal{M}
\label{eq:1}
\end{eqnarray}
where $\Delta_{m} = (\omega_{m}-\omega_{o})$ is the detuning of plasmon mode frequency $\omega_{m}$ from incident field frequency $\omega_{o}$, and $\Delta_{0j} = (\omega_{0j}-\omega_{o})$ the detuning of QE transition frequency $\omega_{0j}$ from the driving source frequency $\omega_{o}$. To induce Stark shift in the excitonic levels, QE is attached to external voltage-bias, as shown in Fig.\ref{fig:1}. Under the second order perturbation treatment, the Stark shift ($\Delta E$) produced in the excited states of QE is referred to as quadratic Stark shift, which is calculated through the relation $\Delta E = -1/2 \alpha E^2$, where E is the applied electric field and $\alpha$ is the polarizability of the atomic system, derived from $9/2(a_o)^3$ with $a_o$ as Bohr radius \cite{Bransden2014}. In our simple model, we ignore electron spin and other effects such as relativistic correction and Lamb shift by considering these effects as small in comparison to applied electric field. In contrast to linear Stark effect, quadratic Stark shift results from the induced dipole moment ($\mu_{qe}$) of the energy states after the application of external voltage-bias. The change in $\mu_{qe}$ is proportional to the strength of the applied electric field. In this way, as the electric field strength increases, the induced dipole moment becomes more significant which causes a pronounced shift in the plexcitonic states. In addition, when the interaction of system with the environment reservoir is taken into account, our plasmon-emitter system acts as an open quantum system under Markovian approximation \cite{Breuer2007}.
\begin{figure*}[t!]
    \centering
    \begin{subfigure}
        \centering
        \includegraphics[height=2in]{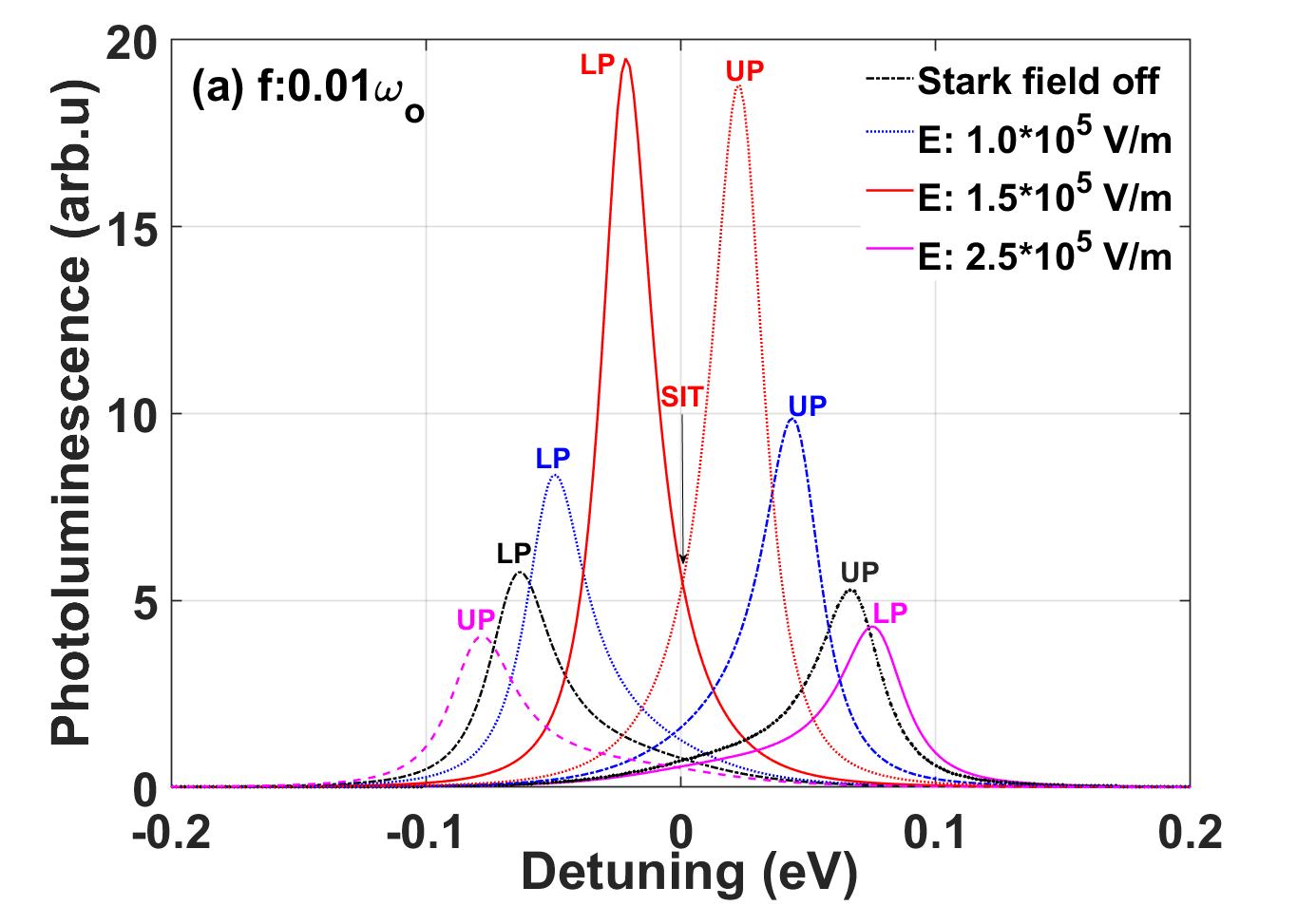}
    \end{subfigure}%
    ~ 
    \begin{subfigure}
        \centering
        \includegraphics[height=2in]{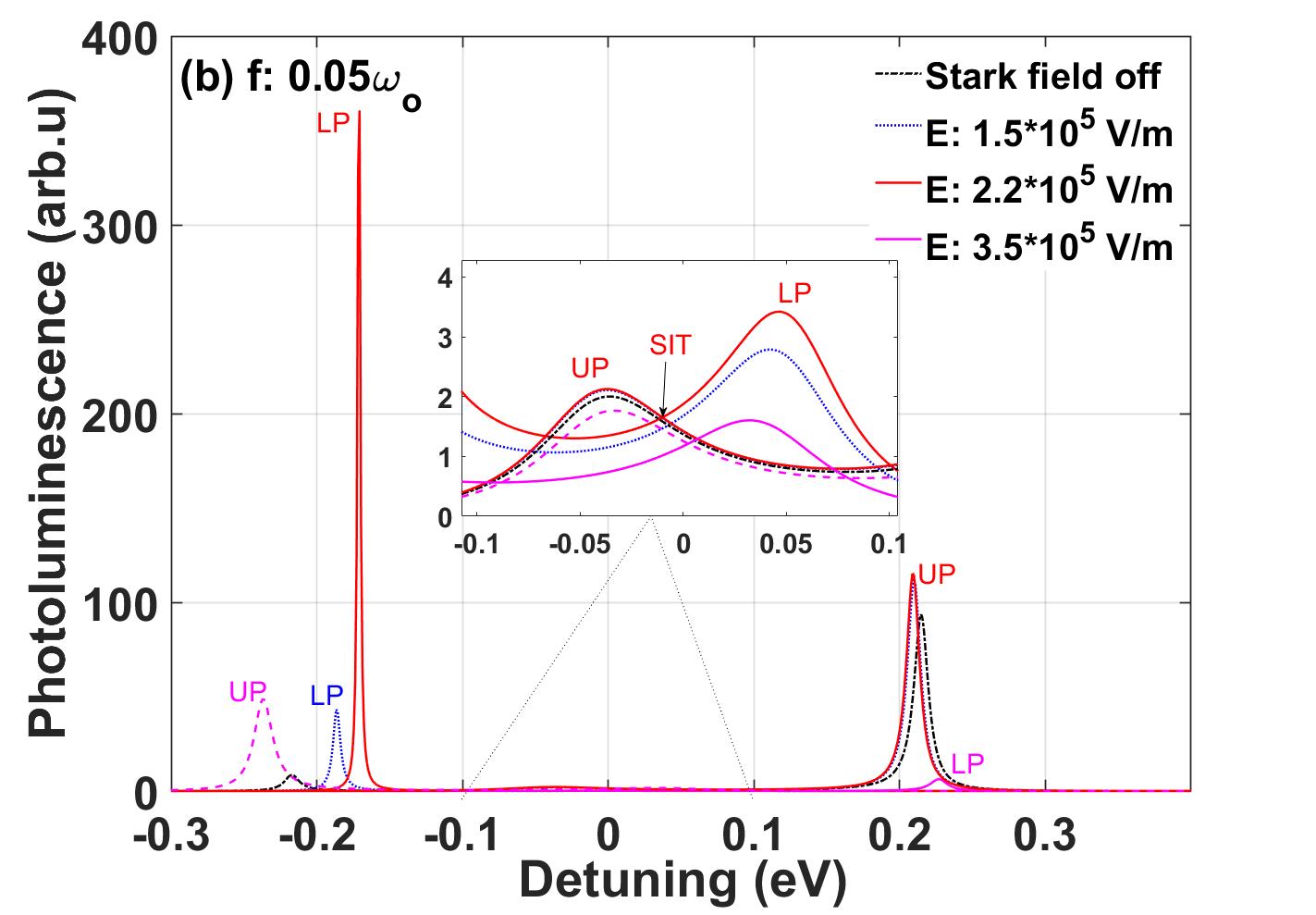}
    \end{subfigure}
    \caption{\label{fig:3} Photoluminescence spectra of Stark induced plexcitonic shifts as a function of detuning $(\Delta_m)$ for (a) weak (f=0.01$\omega_o$) and (b) intermediate (f=0.05$\omega_o$) coupling regimes. The spectral parameters are the same as mentioned in Fig.\ref{fig:2}.}
\end{figure*}
The dynamics of hybrid plasmon-emitter system is evaluated through the Heisenberg-Langevin approach \cite{Waks2010} which provides a simple method to evaluate operators, handle damping and interaction with external field. The equations are as follow.
\begin{eqnarray}
\partial_t\hat{a}= \frac{i}{\hbar}[\hat{\mathcal{H}},\hat{a}]-\frac{\gamma_m}{2}\hat{a}
\label{eq:2}
\end{eqnarray}
\begin{eqnarray}
\partial_t\hat{\sigma}_{0j}= \frac{i}{\hbar}[\hat{\mathcal{H}},\hat{\sigma}_{0j}]-\frac{\gamma_{0j}}{2}\hat{\sigma}_{0j}
\label{eq:3}
\end{eqnarray}
These equations combine the Heisenberg equation of motion with the damping terms resulting from the Markovian interaction with the reservoir which determines the decay rates of plasmon mode ($\gamma_{m}=72$ meV) and the excitonic levels of QE ($\gamma_{0j}$). The equations of motion derived for plasmon mode amplitude $\langle\hat{a}\rangle$ and off-diagonal density matrix elements 
$\langle\hat{\sigma}_{0j}\rangle$ of QE using bosonic commutation relations are as follow.
\begin{equation}
\partial_t\langle\hat{a}\rangle= -[i(\Delta_m+\gamma_m/2)]\langle\hat{a}\rangle +f\langle\hat{\sigma}_{01}\rangle+ f\langle\hat{\sigma}_{02}\rangle+ \mathcal{M}
\label{eq:4}
\end{equation}
\begin{equation}
 \partial_t\langle\hat{\sigma}_{01}\rangle= -[i(\Delta_{01}+\Delta E)+\gamma_{01}/2]\langle\hat{\sigma}_{01}\rangle + f\langle\hat{a}\rangle 
 \label{eq:5}
\end{equation}
\begin{equation}
\partial_t\langle\hat{\sigma}_{02}\rangle= -[i(\Delta_{02}-\Delta E)+\gamma_{02}/2]\langle\hat{\sigma}_{02}\rangle + f\langle\hat{a}\rangle
 \label{eq:6}
\end{equation}

We deliberately select a smaller amplitudes for probe field. This choice offers the additional advantage of minimizing other nonlinear effects, thus enhancing the precision of our approach in the weak field limit. Therefore, while driving the analytical solutions, we employ weak field approximation by considering the intensity of incident field is sufficiently weak to demonstrate our ability to perform Stark tuning even at low intensities. In this way, we evaluate the linear dynamics of the system and disregard higher-order terms and hence, we also ignore the noise operator terms. Moreover, in the weak field limit, the excitonic population $\langle\hat{\sigma}_{0j}^\dagger\hat{\sigma}_{0j}\rangle \ll 1$ is minute, therefore, we have $\langle\hat{\sigma}_{00}\rangle = 1$ and $\langle\hat{\sigma}_{jj}\rangle = 0$ \cite{Li2021}. We perform the time evolution of Eq.\ref{eq:4}-\ref{eq:6} numerically through Runge-Kutta method using Matlab program and compute the output scattered intensity of hybrid plasmon-emitter system given by the relation, $I_{sca}=\vert \langle\hat{\sigma}_{0j}\rangle+\langle\hat{a}\rangle \vert^2$ \cite{Ridolfo2010, Waks2010}.\\

\begin{figure*}[t!]
    \centering
    \begin{subfigure}
        \centering
        \includegraphics[height=2in]{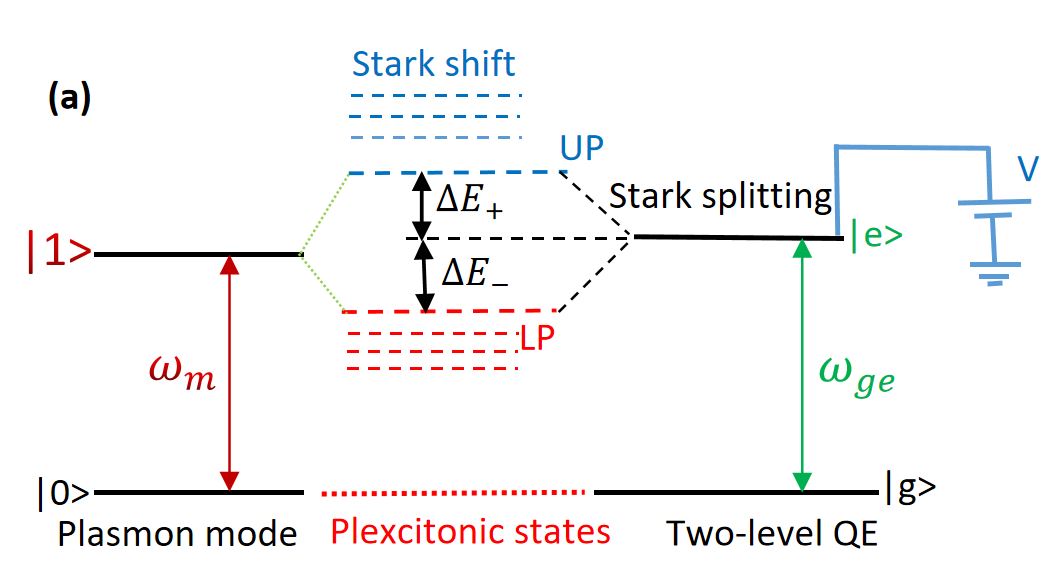}
    \end{subfigure}\\%
    ~ 
    \begin{subfigure}
        \centering
        \includegraphics[height=2in]{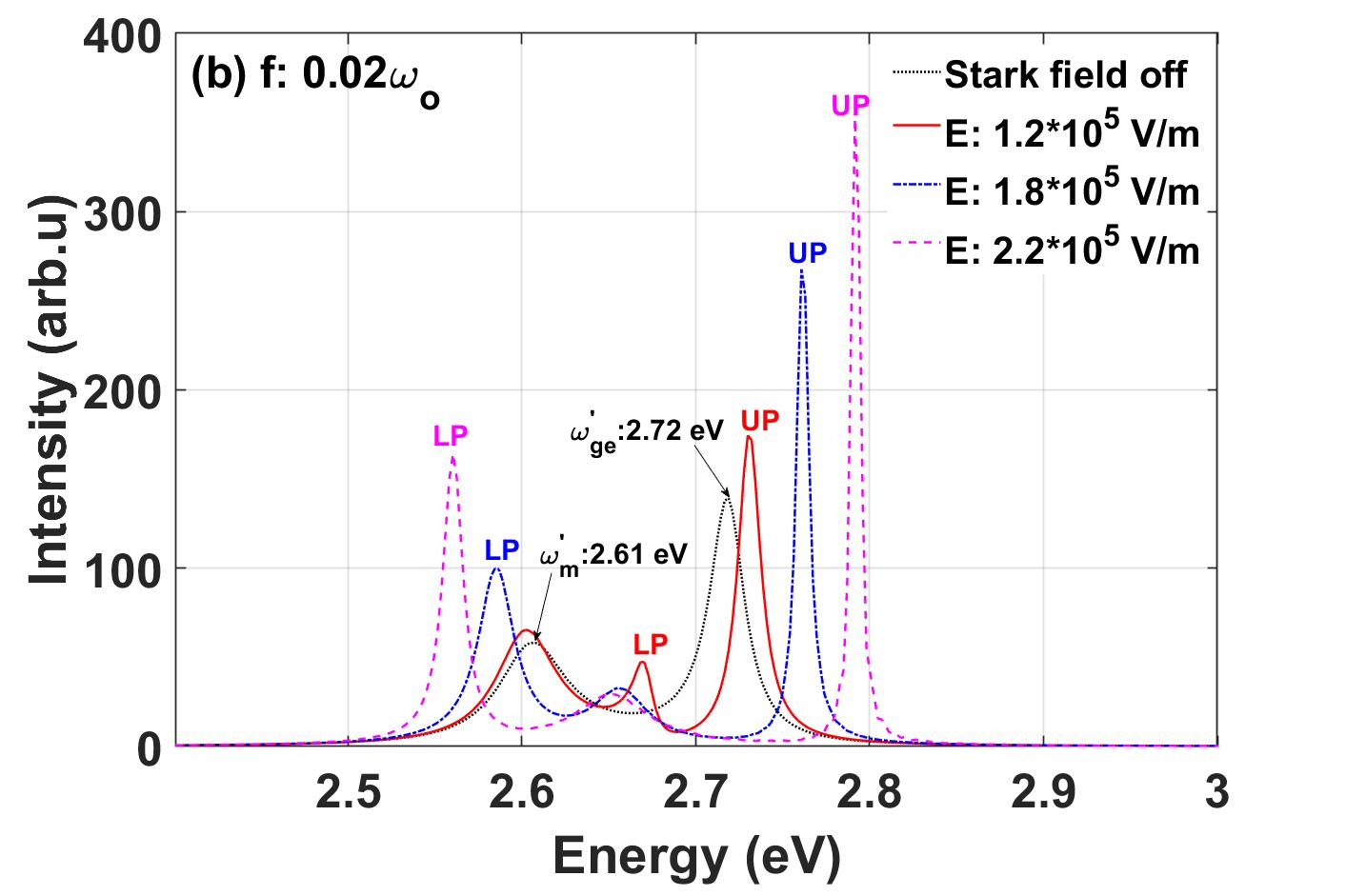}
    \end{subfigure}
		 ~ 
    \begin{subfigure}
        \centering
        \includegraphics[height=2in]{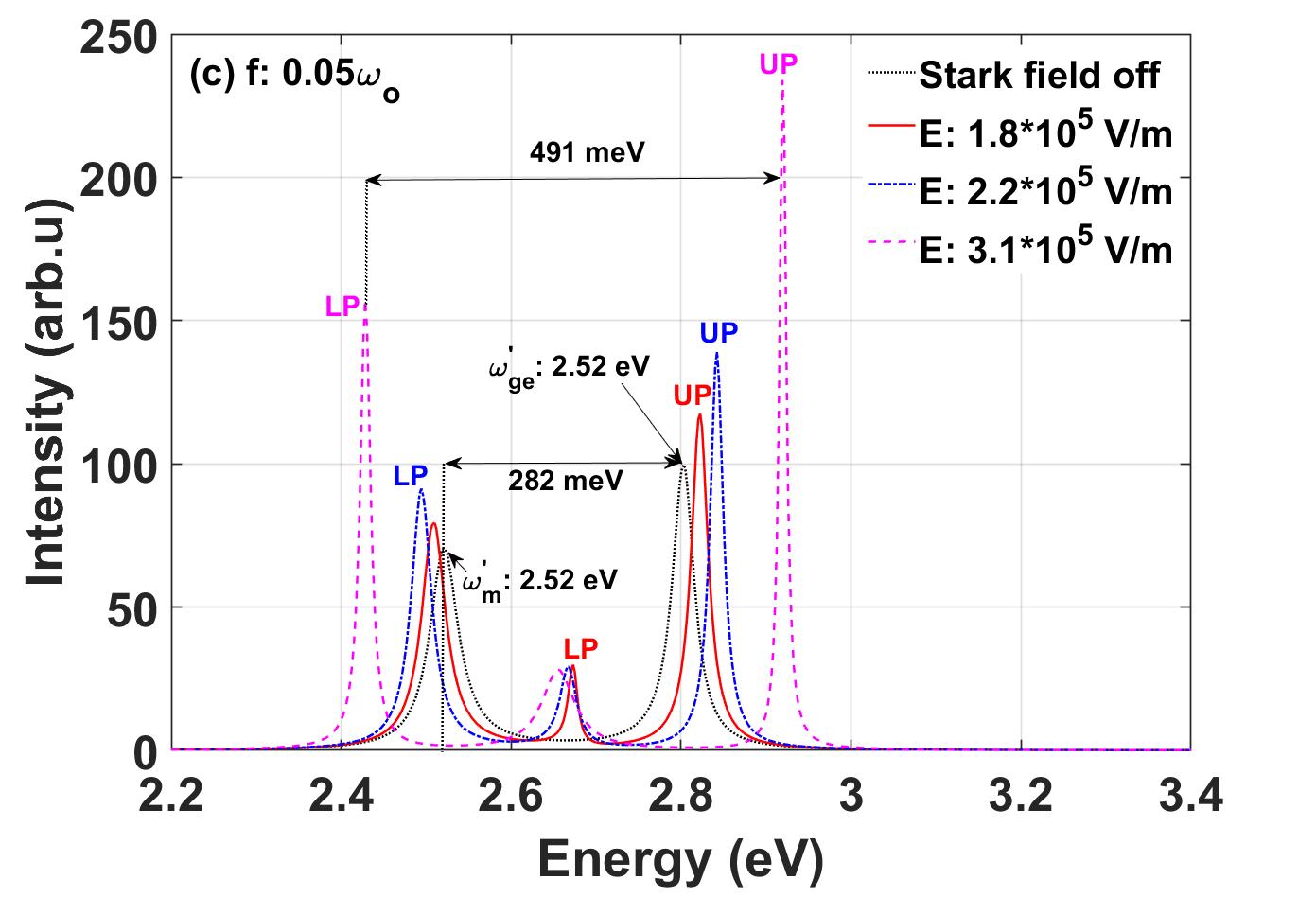}
    \end{subfigure}
    \caption{\label{fig:4} (a) Schematic diagram of PM coupled to two-level QE $(\omega_m\approx\omega_{01})$, and the tuning of plexcitonic states through Stark induced splitting and shifts. Stark tuned plexcitonic states and double Fano resonance in the resonantly coupled plasmon-emitter system for (b) weak and (c) intermediate coupling.}
\end{figure*}

\subsection{Stark induced shift in off-resonantly coupled plexcitonic states}
We analyze the scattered intensity of plasmon-emitter system in the weak, intermediate and strong coupling regimes \cite{Leng2018}, as shown in Fig.\ref{fig:2}. The energy spectra show three plexcitonic states corresponding to Lower Plexciton (LP), Plasmon Mode (PM), and Upper Plexciton (UP) in the hybrid system. The transition frequencies of QE $\omega_{01}= 2.5$ eV and $\omega_{02}= 2.7$, eV are kept off-resonant to PM frequency, $\omega_{m}= 2.64$ eV. The driving field is also taken off-resonant from both QE and PM, with excitation frequency, $\omega_{0}= 1.997$ eV. In the absence of Stark field, a shift is observed in the excitonic levels due to coupling of bare states with plasmon mode, such type of non-resonant shift results in the bending of the bare energy levels. In the dispersive limit, when $\Delta= (\omega_{0j}-\omega_m) \gg f$ and $f\ll (\gamma_{0j}-\gamma_{m})/2$, the shift in the plexcitonic states occurs due to Plasmonic Stark Effect (PSE) \cite{Zheng2021}. In Fig.\ref{fig:2}(a), we observe PSE shift in the excitonic levels forming UP and LP states with energies $2.72$ eV and $2.55$ eV, respectively. When the Stark field is turned on, UP undergoes a redshift, while LP blue shifts towards plasmon resonance frequency. Upon increasing the electric field strength, the incoherent plexcitonic states enter the resonant regime resulting in the path interference of UP and LP states with PM [see red curve in Figures.\ref{fig:2}(a)-(b)]. The coherent interaction of plexcitons due to Stark field yields Fano resonance \cite{Wu2010, Bayrakli2021, Artvin_2020} and a transparency window is appeared at the plasmon resonance frequency ($\omega_m= 2.64$ eV). We named this transparency as Stark induced transparency (SIT).\\
A significant increase in the quadratic Stark shift is observed with the increase in the electric field strength which results in a significant red- and blue-shift of UP and LP states from the bare energy levels, respectively. In the intermediate coupling regime ($f= 0.05\omega_o$), the new hybrid states are UP $= 2.79$ eV, LP $= 2.44$ eV and PM at $\omega_{m}^{'}$ $= 2.62$ eV. When the Stark field is turned on the UP level red shifts to lower energy level with a decrease in the energy of 26 meV and LP blue-shifts to a higher level with an increase in the energy up to 39 meV. As the levels transit towards the resonant energy state of polaritons, the coherent interaction yields Fano resonance along with transparency at $2.60$ eV. The red and blue shifts in the UP and LP states change the overall dynamics of the hybrid molecular system. The crossing of plexciton states to either side leads to vacuum Rabi splitting due to a quadratic shift in LP mode with an energy difference of 226 meV from polariton mode resonance and the shift in UP cause splitting of 194 meV. A small increase in the electric field transforms incoherent states to a resonantly oscillating coherent system. Similar effects are observed in the strongly coupled plexcitonic states, in which Stark field induces maximum Rabi splitting of $\Omega\le 350$ meV. Here, $\Omega$ is the difference between transition frequencies of shifted levels with corresponding polariton mode frequency ($\omega_{m}^{'}$).\\   
To validate these shifts, we plot the photoluminescence (PL) spectra as a function of detuning ($\Delta_m$). The photoluminescence for a cavity-emitter coupled system is defined as \cite{Cui2006}, 
\begin{equation}
\mathrm{PL}= \frac{\gamma}{\pi}\bigg\vert\frac{-i f}{(\gamma_{+}+i(\Delta/2-i\Delta_m)^2+(\Omega_{R}/2)^2}\bigg\vert^2  
 \label{eq:7}
\end{equation}
where $\Omega_R$ is the complex form of generalized Rabi frequency defined as, $\Omega_R=\sqrt{4f^2-(4\gamma_{-}+i\Delta_{0j})^2}$ with the energy dissipation rates, $\gamma_{+}=\vert\frac{\gamma_{oj}+\gamma_{m}}{4}\vert$ and $\gamma_{-}=\vert\frac{\gamma_{oj}-\gamma_{m}}{4}\vert$. The detuning between shifted energies of plexcitons states and plasmon mode is defined as $\Delta = (\omega_{0j}^{'} - \omega_m^{'}$). We calculate PL intensity for different values of Stark field and evaluate the quadratic shift in UP and LP for three different coupling regimes as shown in Fig.\ref{fig:3}. In the dissipative regime, the plexciton with nearly resonant states close to zero detuning region [point of SIT in Fig.\ref{fig:3}(a)] yields maximum PL. Whereas, in Fig.\ref{fig:3}(b), the UP and LP peaks split with large detuning and maximum transitions occur at the point of crossing near zero detuning [see inset Fig.\ref{fig:3}(b)] and as the levels move away towards large detuning, the PL from both UP and LP states decreases profoundly (purple dashed and solid curve). We also obtained similar results for the strong coupling regime. The splitting of UP/LP states enhances further with increase in field strength and maximum PL occur at the induced transparency (the results are not shown here). The tuning of plexcitonic levels not only produces a spectral shift but it also switches the photoluminescence intensity to on/off by modulating the coherent dynamics of hybrid system through external probe (see Fig.\ref{fig:3}). Moreover, our proposed system demonstrates the tuning of PL in the optical frequency range. In this way, Stark tuning of an incoherent quantum system transforms it into an actively tunable coherent photonic device that carries strong potential for nanoscale lasing and on-demand PIC technologies.\\

\subsection{Stark induced splitting in resonantly coupled plexcitonic states}
In this section, we illustrate the Stark splitting in a two-level QE through external probe field which induces double Fano resonance and Rabi splitting in the resonantly coupled plasmon-emitter system. For this, we use the same Hamiltonian as defined in Eq.\ref{eq:1} by replacing $\Delta_{01}$ and $\Delta_{02}$ with detuning parameter ($\Delta_{q}$) and derive the equations of motion for plasmon mode and off-diagonal density matrix elements of two-level QE by using Eq.\ref{eq:2} and Eq.\ref{eq:3} as follow,

\begin{equation}
\partial_t\langle\hat{a}\rangle= -[i(\Delta_m+\gamma_m/2)]\langle\hat{a}\rangle +f\langle\hat{\sigma}_{01}\rangle+ f\langle\hat{\sigma}_{02}\rangle+ \mathcal{M}
 \label{eq:8}
\end{equation}
\begin{equation}
 \partial_t\langle\hat{\sigma}_{01}\rangle= -[i(\Delta_{q}+\Delta E_{-})+\gamma_{01}/2]\langle\hat{\sigma}_{01}\rangle + f\langle\hat{a}\rangle 
 \label{eq:9}
\end{equation}
\begin{equation}
\partial_t\langle\hat{\sigma}_{02}\rangle= -[i(\Delta_{q}-\Delta E_{+})+\gamma_{02}/2]\langle\hat{\sigma}_{02}\rangle + f\langle\hat{a}\rangle
 \label{eq:10}
\end{equation}
where the probe field splits the excited state of QE with transition frequency ($\omega_{ge}$) and the shift in the levels is evaluated as $\Delta E$ (see Fig.\ref{fig:4}). $\Delta_{q}$ = ($\omega_{ge}$ - $\omega_{o}$) is the detuning of two-level QE transition frequency ($\omega_{ge}$) from the incident light frequency ($\omega_o$). For the off-resonant excitation, the frequency of pump field is taken as $\omega_o= 1.997$ eV. Nevertheless, plasmon mode and QE couple resonantly with frequencies, $\omega_m= 2.64$ eV and $\omega_{eg}= 2.68$ eV, respectively. 

\begin{figure}[t!]
\includegraphics[scale=0.17]{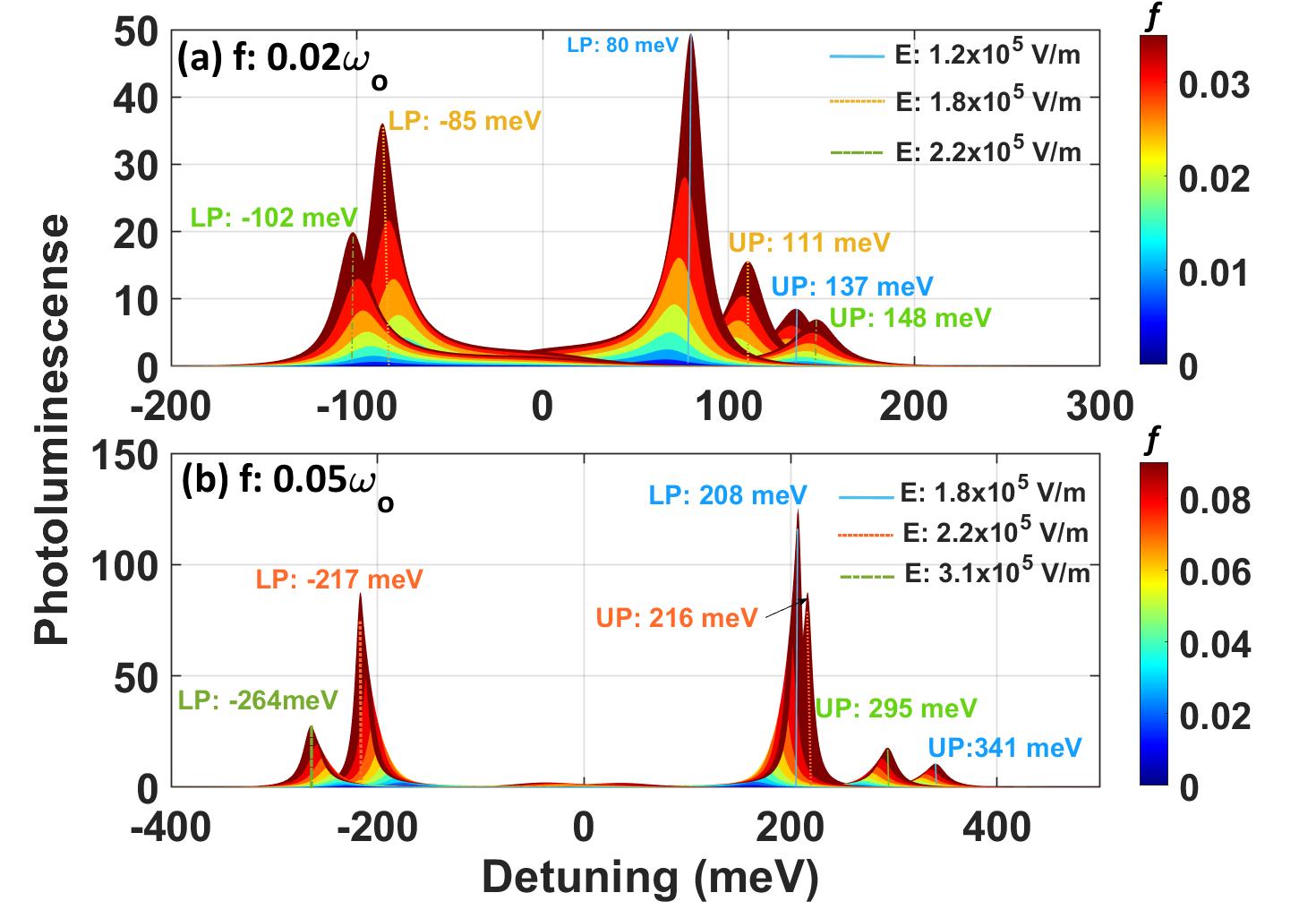}
\caption{\label{fig:5} Photoluminescence spectra of Stark tuned plexcitonic states as a function of detuning $(\Delta_m)$ for (a) $f=0.02\omega_o$. (b) $f=0.05\omega_o$. UP plexciton blue-shifts and LP redshifts as the Stark field strength increases.}
\end{figure}

Whereas the values of $\gamma_m$, $\gamma_{0j}$ and $\mathcal{M}$ are the same as defined in Fig.\ref{fig:2}. After performing the time-evolution of Eq.\ref{eq:8}-\ref{eq:10} numerically, we compute the scattered intensity $(I_{sca})$ of hybrid system as a function of excitation wavelength. Fig.\ref{fig:4}(a) shows the schematic of energy-level diagram of plexcitonic states, plasmon mode and a two-level QE. We calculate the energy spectra of plexcitonic states with Stark splitting and shift for different values of electric field. The black dotted curve in Fig.\ref{fig:4}(b) and Fig.\ref{fig:4}(c) indicate two peaks which correspond to hybrid modes in the absence of probe field. The peak spectral positions of hybrid modes appear at $\omega_m^{'} = 2.61$ eV and $\omega_{ge}^{'}  = 2.72$ eV for the weakly coupled system  ($f= 0.02\omega_o$) and $2.52$ eV and $2.8$ eV for intermediate coupling ($f= 0.05\omega_o$), respectively. When the Stark field is turned on the excitonic level of QE splits into two which then couple to existing polaritonic mode. The coupling of discrete states with continuum mode results in the path interference of three coherent states giving rise to double Fano resonance. The new plexcitonic modes show three peaks with energies $2.52$ eV, $2.67$ eV and $2.73$ eV for PM, LP and UP state, respectively [see red in Fig.\ref{fig:4}(b) and \ref{fig:4}(c)].
As the field strength increases the plexcitonic states indicate the signature of Mollow triplets \cite{Ge2013}. Though in our case, Mollow triplets result from the interaction of intense LSP with Stark split excitonic levels. With the increase in the probe field, the LP red-shifts and UP blue-shifts with large detuning with PM. For $f=0.05\omega_o$ coupling, hybrid mode split with an energy shift of 282 meV. In the presence of Stark field, the splitting generates two new plexcitonic states with energies $2.7$ eV (UP) and $2.8$ eV (LP). As the field strength increases to $3.1\times10^5$ V/m, the plexciton dynamics drastically change with a maximum energy shift of 491 meV, [Fig.\ref{fig:4}(c)].\\
We validate the plexciton tuning through photoluminescence spectra and evaluate the optical response of the system for different values of Stark field. Fig.\ref{fig:5}(a) and \ref{fig:5}(b) shows the contour plots of PL for UP and LP states as a function of detuning ($\Delta_m$). For the value of $E= 1.2\times10^5 V/m$ and $E= 1.8\times10^5 V/m$, LP and UP peaks show the signature of double Fano resonance, with the minimum energy difference between plexciton (a) 57 meV and (b) 133 meV, respectively. As the field strength increases, the spectral energy and PL intensity of UP and LP are modulated in a profound manner. The UP plexciton blue shifts with maximum energy of 148 meV and LP redshifts to 102 meV [Fig.\ref{fig:5}(a)]. In contrast to this, for strongly coupled LP and UP modes, the spectral energy shifts are more pronounced with detuning 264 meV and 341 meV, respectively, see Fig.\ref{fig:5}(b). Here, some small peaks also appear close to zero detuning along with intense UP/LP peaks which shows the signature of Rabi splitting \cite{Leng2018} in plexciton. Nevertheless, as the splitting between UP and LP states increases, the PL intensity decreases gradually indicating the switching of spontaneous photon emission from high (on) to low (off). In this way, Stark splitting of exciton not only modulates the energy of hybrid states but also actively tunes the photoluminescence intensity (on/off) as a function of the probe field. 

\section{CONCLUSIONS} 
In summary, we theoretically investigate coherent control of plexcitonic states in incoherent quantum systems through optical Stark effect (OSE). We demonstrate Stark tuning of hybrid plexciton modes in a coupled plasmon-emitter system. A small perturbation in the applied field modulates the energy eigenstates of quantum emitter which modifies the hybrid modes in a coherent manner. At first, we evaluate the Stark-induced spectral shifts of plexciton states in an off-resonant coupled system. The pronounced resonant shifts in the excitonic levels result in the coherent interference of dressed states leading to tunable Fano resonance. We also report the Stark induced transparency in hybrid states in both weak and intermediate coupling regimes. Furthermore, we investigated double Fano resonance in resonantly coupled plexciton modes due to splitting of eigen energy state of QE. The Stark-induced splitting shows the signature of Mollow triplets in the plexcitonic modes with maximum energy splitting upto 491 meV between upper and lower plexcitons. We also explore the impact of Stark tuning of plexcitons on photoluminescence spectra, which shows active control of photon transitions and on/off switching of spontaneous emission in the visible regime. Our proposed coherent optical control of quantum states through Stark effect clearly indicates its potential in quantum information processing and quantum computing applications. In addition to this, such tunable signatures can be used for engineering dynamical nanophotonic systems with potential applications such as lasing, ultrafast switching, optical modulation, single-photon emission and sensing applications. 
 
\begin{acknowledgments}
R.S. and H.A. acknowledge support from TUBITAK No. 121F030 and 123F156.
\end{acknowledgments}


\providecommand{\noopsort}[1]{}\providecommand{\singleletter}[1]{#1}%

\end{document}